# A Symmetric Superconducting Dome Hosts Non-Fermi Liquid Behavior at Optimal Doping in $MoS_2$

Qiao Chen[1, 2, †], Chengyu Yan[1, 3, †, *], Dino Novko[4], Changshuai Lan[1], Huiqin Jian[5], Yi Yan[1], Xinming Zhao[1], Yihang Li[1], Huai Guan[1], Bo Gao[6, 7], Zhong Wan[2, 8], Shun Wang[1, 3, *]

[1]MOE Key Laboratory of Fundamental Physical Quantities Measurement & Hubei Key Laboratory of Gravitation and Quantum Physics, National Gravimetry Laboratory and School of Physics, Huazhong University of Science and Technology, Wuhan 430074, China

[2]Quantum Science Center of Guangdong-Hong Kong-Macao Greater Bay Area, Shenzhen 518045, China

[3]Institute for Quantum Science and Engineering, Huazhong University of Science and Technology, Wuhan 430074, China

[4]Centre for Advanced Laser Techniques, Institute of Physics, Zagreb 10000, Croatia.

[5]College of Science, Guizhou Institute of Technology, Guiyang 550025, China

[6]State Key Laboratory of Micro-nano Engineering Science, Tsung-Dao Lee Institute, Shanghai Jiao Tong University, Shanghai 201210, China

[7]Shanghai Research Center for Quantum Sciences, Shanghai 201315, China

[8]Department of Physics, The University of Hong Kong, Pokfulam Road, Hong Kong 999077, China

Q. Chen and C. Yan contribute equally.

Corresponding Authors: Chengyu Yan (chengyu_yan@hust.edu.cn); Shun Wang (shun@hust.edu.cn)

## Abstract

The similarities between the phase diagrams of ionic liquid-gated transition metal dichalcogenides (TMDCs) and high-temperature superconductors have garnered considerable interest due to the presence of a superconducting dome with a non-monotonic dependence of the superconducting order parameter as a function of charge carrier density. However, the lack of a complete superconducting dome and insights into the normal state in ionic liquid-gated TMDCs prevents a detailed mapping between the two systems. In this work, we obtain a symmetric superconducting dome that extends from deep underdoped regime all the way to deep overdoped regime in ionic liquid gated $MoS_2$ with a refined gating protocol. We demonstrate that the dome is anticorrelated with the evolution of non-Fermi liquid behavior in the normal states. The scattering rate in the non-Fermi liquid regime can reach Planckian limit. The results may shed light on the emergence of superconductivity in TMDCs.

## Introduction

Despite its simplicity in terms of crystal structure, transition metal dichalcogenides (TMDCs) offer a versatile platform to explore two-dimensional superconductivity[1-3], superconducting dome[4-7], charge density waves[8-10], and non-Fermi liquid[11, 12]. It has been widely marked that the phase diagram of TMDCs strikingly mimics that of high-$T_c$ superconductors such as cuprates[13-16] and iron-based superconductors[17-20]. It is therefore proposed that a complete phase diagram of TMDCs, fully elucidate both underdoped and doverdoped region by encompassing a wide doping range, not only sheds light on the intriguing properties of TMDCs but also high-$T_c$ superconductors.

Ionic liquid gating has been a leading method in the survey of the phase diagram of TMDCs[4-7] thanks to its extraordinary capability in tuning the doping level. Taking $MoS_2$ as an example, the ionic liquid gating method can inject carriers into $MoS_2$ via electrostatic doping, enabling continuous tuning from an insulating to a metallic and finally a superconducting state[4, 5, 21]. However, the bulk of previous works in ionic liquid gated TMDCs only manage to map the underdoped regime and the vicinity of the optimal doping, while little progress has been accomplished in the overdoped regime of the superconducting dome[22], leaving an unexplored region in the superconducting phase diagram. Moreover, little attention has been paid to the normal states of ionic liquid-gated TMDCs. This stands in sharp contrast to high-$T_c$ superconductors. The absence of a complete superconducting dome and insights into the normal state hinder the comprehensive understanding the emergence of superconductivity in TMDCs.

In this work, we obtain a complete superconducting dome in ionic liquid-gated $MoS_2$ flakes by adapting a refined gating protocol. The refined gating protocol unveils that the highest achievable carrier concentration actually occurs at moderate gate voltage in an etched TMDC Hall bar. The carrier concentration drops at large gate voltage due to the enhanced localization effect[23, 24]. The superconducting dome is highly symmetric against carrier concentration, with critical temperature $T_c$ following a square root dependence in both underdoped and overdoped regime. More interestingly, we find that the superconducting dome is anticorrelated with the evolution of non-Fermi liquid behavior in the normal states, where the non-Fermi liquid region occurs in the vicinity of optimal doping and gets flanked by a Fermi liquid region on both sides. The scattering rate in the non-Fermi liquid regime can reach Planckian limit. These results suggest phase diagram in ionic liquid-gated TMDCs may go beyond the electron-phonon interaction scenario and prompt a more detailed cross-check between TMDCs and high-$T_c$ superconductors.

## Result

### Symmetric superconducting dome

This study employs $MoS_2$ flakes (~10 nm in thickness), etched into standard Hall bar shape (see Method and SI section 1 for the details of fabrication) and gated with ionic liquid DEME-TFSI (N,N-diethyl-N-methyl-N-(2-methoxyethyl) ammonium bis (trifluoromethylsulfonyl)-imide) for electrical transport measurements, as shown in Fig. 1(a). The ionic liquid can accumulate carriers at the ionic liquid-$MoS_2$ interface and drives the formation of a superconducting state within a certain range of carrier concentration[1, 2, 4]. Since it requires a high carrier concentration to explore the overdoped region of the superconducting dome, we first determined the highest achievable carrier concentration. It was noticed that the carrier concentration followed a non-monotonic behavior at both 220 K (where the ionic liquid was not frozen) and 15 K (where the ionic liquid was already frozen) against gate voltage $V_{ig}$ as depicted in Fig. 1(b), with a maximal $n_{2D}$ of the device of 17 × $10^{13}$ $cm^{-2}$ that should noticeably intrude into the overdoped region. The $n_{2D}$ is determined by Hall effect measurement, as shown in Fig. S2. We demonstrated that the non-monotonic behavior was a consequence of the localization of carriers due to the formation of zig-zag distributed ions at large $V_{ig}$ instead of electrochemical reaction in our previous work[23]. The shift of maxima of $n_{2D}$ at 15 K can be qualitatively explained with the localization framework: Lowering temperature will suppress thermal activation and thus enhanced the effective localization strength, however, this effect is partially offset by the increased carrier mobility which makes it less efficient to capture a mobile charge carrier in the first place. The exact direction of the shift depends on the interplay of the two effects, and may be different across different TMDCs. Further lowering the temperature to the superconducting regime would not change the $n_{2D}$-$V_{ig}$ relation significantly, hence we used $n_{2D}$ determined at 15 K to quantify carrier concentration in low temperature regime. In the following study, we kept $V_{ig}$ below 4.5 V to avoid the localization effect becoming dominant.

When $V_{ig}$ was set between 2.5 and 4.5 V, the device exhibited a superconducting phase with an enhanced in-plane upper critical field, consistent with previous works[1, 2, 25]. Fig. 1(c)&(d) summarizes the typical behavior at $n_{2D}$=12.1×$10^{13}$ $cm^{-2}$ (see SI section 2 for the details). Similar measurements at other carrier concentrations (or $V_{ig}$) were summarized in Fig. 2(a). It was noticed that superconducting critical temperature $T_c$ (defined at 90% $R_n$) initially increased with rising $n_{2D}$ up to 12.1 × $10^{13}$ $cm^{-2}$, and then dropped as $n_{2D}$ further increases. Meanwhile, the normal state resistance decreased monotonically with increasing $n_{2D}$, in contrast with the re-entrant insulating behavior observed in 1L-$WS_2$[6, 24].

By plotting the extracted $T_c$ as a function of $n_{2D}$, as shown in Fig. 2(b), a complete superconducting dome covering both the underdoped and overdoped regimes is fully revealed, we note that most of the previous works only managed to capture the underdoped regime and slightly penetrated into the overdoped side[4, 5] in gated $MoS_2$. We stressed that the measured $T_c$ in the underdoped side was in good agreement with previous works[4, 5] (see SI section 3 for the details). It was then necessary to analyze the result more quantitatively by comparing the overdoped regime with the underdoped regime. It was interesting to note that $T_c \propto |n_{2D} - n_c|^{0.52\pm0.05}$ in both underdoped and overdoped regime, with $n_c = 6.57 \times 10^{13}$ cm$^{-2}$ on the underdoped side and $n_c = 17.91 \times 10^{13}$ cm$^{-2}$ on the overdoped side, suggesting a symmetric superconducting dome. This observation was in staggering contrast with previous measurement[22], where a highly asymmetric dome with a long tail in the overdoped side was reported. We speculated the difference might be associated with the details of gating (see discussion in SI section 3).

The symmetric dome and the square root dependence of $T_c$ on $n_{2D}$ strikingly resembled that in unconventional superconductors such as $TiSe_2$, cuprate and iron-based superconductors[8, 17, 26-28]. This resemblance can be further appreciated by evaluating the ratio $T_c/T_F$, where $T_c$ is the superconducting transition temperature, and $T_F$ is the Fermi temperature (see SI section 4 for the details). Again, $T_c/T_F \approx 0.006$ at the optimal doping fell in the regime typical for cuprate, iron-based, heavy-fermion, and twisted-2D systems. This highlights similarities between gated $MoS_2$ and other unconventional superconductors.

**Non-Fermi liquid behavior in the normal state**

Commonly, unconventional superconductors exhibit exotic quantum phase in the normal state[12-19, 29-31]. It is therefore important to understand the normal state behaviour to gain more insight into the mystery of the superconducting phase.

First of all, we measured the temperature dependence of resistance (*RT* curve) in normal state in the dome region as shown in Fig. 3(a). It was found that the behavior can be satisfactorily captured by $R = R_0 + A \times T^{\alpha}$, where $R_0$ is the residual resistance and $A$ is a system specific parameter to be determined. We did not go above $T$=170 K in the measurement because the ionic liquid will begin to melt at the point, *RT* curve may be destabilized due to the reconfiguration of the distribution of ions, which may lead to an unreliable interpretation of the results. Surprisingly, $\alpha$ exhibits a saddle shape with respect to $n_{2D}$, which was anticorrelated with the evolution of $T_c$, as highlighted in Fig. 3(b) (see SI section 5 for more details). The anticorrelation was widely reported in unconventional superconductors like $TiSe_2$, cuprates, iron-based, and nickelates superconductors[8, 17, 20, 26, 28, 32, 33]. More specifically, $\alpha$ approached 2 at the edge of the dome, a typical result of Fermi liquid (FL) behavior; it approached 1 at the optimal doping. We will verify shortly that the linear $RT$ curve was associated with non-Fermi liquid (non-FL), even an indication of strange metal. It is also interesting to mark that the $RT$ curve tends to be saturated at the highest doping at ~160 K, which could be related to bad metal behavior[11] or charge density wave (CDW)[34, 35] and warranted further study.

The observation of a linear $RT$ curve alone cannot sufficient to claim non-FL behavior. The nature of the linear $RT$ curve should be carefully examined with the help of the scattering rate.

First of all, a strong anisotropy in the longitudinal and transverse scattering rate at the optimal doping was observed. Typically, this analysis was performed in high-$T_c$ superconductors by analyzing the difference in the temperature dependence of longitudinal resistivity ($\rho_{xx}$) and Hall angle $\Theta_H$ = arctan ($\sigma_{xy}/\sigma_{xx}$), where $\sigma_{xy}$ and $\sigma_{xx}$ are the Hall and longitudinal conductivities, respectively. We adapted the same approach in our system. It was shown in Fig. 4(a) that $\rho_{xx}$ at zero magnetic field followed a linear dependence, while $\Theta_H$ followed a quadratic dependence on temperature (see SI Section 6 for the raw data). This observation is in line with reports in high-$T_c$ superconductors[18, 31, 36-38]. Strictly speaking, the approach based on the comparison between longitudinal resistivity and Hall angle only applies for a single parabolic band. This is should the case in our device, where a linear Hall resistance without any plateaus or turning points are observed (Fig. S2). Nevertheless, to avoid the scenario that we missed the contribution of multiple bands due to the limitation of available field, we further supplemented the argument in the light of extended Kohler's rule[39, 40] should multiple bands contribute to the transport. The original Kohler's rule stated that the magnetoresistance ($MR$) should exhibited a scaling behavior, $MR = \frac{R_{xx}-R_0}{R_0} = f(\gamma \frac{B}{\rho_0})$, where $f$ is a system-specific scaling functionality, $\rho_0$ is the zero-field resistivity at a given temperature, and $\gamma$ is related to carrier concentration and effective mass. The temperature dependence of scattering rate was encoded into $\rho_0$. The detailed scaling of MR can be found in section 6 of SI[41] , which mimics that in high $T_c$ superconductor. Recent theoretical works[40] pointed out, if the transverse transport shares the same scattering rate with the longitudinal transport, then the Hall resistance $R_{xy}$ should also follow a scaling $\frac{R_{xy}}{R_0} = g(\gamma \frac{B}{\rho_0})$, where $g$ is a system-specific scaling functionality that was generally different from $f$. Of course, in the special case of a single parabolic band, $g\left(\gamma\frac{B}{\rho_0}\right) \propto \gamma\frac{B}{\rho_0}$, then the scaling essentially recovers the result $\frac{R_{xy}}{B} \propto \frac{1}{n}$. The central information is that the same set of scaling parameters for *MR* should lead to the scaling of $R_{xy}$. However, we found that $R_{xy}$ measured at different temperatures could not collapse by plugging the scaling parameters determined by *MR*, as enclosed in Fig. 4(b). This discrepancy strongly suggested a difference in the longitudinal and transverse scattering rate regardless of the number of bands, a hallmark of non-FL behavior.

Second, we demonstrated that the zero-field scattering rate can reach the Planckian limit. The scattering rate can be obtained from the Drude relation $\rho = \frac{m^*\Gamma}{ne^2}$, where $\rho$ is the resistivity, $m^*$ is the effective mass, $\Gamma$ is the transport scattering rate, $n$ is the carrier density, and $e$ is the elementary charge. The Planckian dissipation limit is given by $\Gamma_{\mathrm{p}} = \frac{k_{\mathrm{B}}T}{\hbar}$, where $k_{\mathrm{B}}$ is the Boltzmann constant, $T$ is the temperature, and $\hbar$ is the reduced Planck constant. The dimensionless ratio $C$, which compares the scattering rate to the Planckian scale, is therefore $C = \frac{\Gamma}{\Gamma_{\mathrm{p}}} = \frac{\hbar ne^2}{k_{\mathrm{B}}m^*}A_{\square}$, where $A_{\square}$ is the slope of the sheet-resistivity, extracted by the linear *RT* curve. Here we have assumed that the effective mass is temperature independent within the temperature range. This assertion stands if phonon is the major contribution for mass renormalization. In the phonon dictated scenario, mass renormalization typically starts kicking in at $T_{\mathrm{ph}}/3$ where $T_{\mathrm{ph}} = E_{\mathrm{ph}}/k_{\mathrm{B}}$ and $E_{\mathrm{ph}}$ is the phonon energy[42]. In $MoS_2$, $T_{\mathrm{ph}}/3$ is around 150 K according to phonon energy measured by H. Tornatzky et al.[43], almost the upper bond of the temperature of interest in the measurement. Using $m^*=1.1m_0$ of electron-doped $MoS_2$[44] and the carrier density $n$ extracted from Hall measurements, we find that $C \approx 1.2$ in the optimally doped region in two devices (see SI section 7 for more details) as shown in Fig. 4(c). This indicated that the scattering rate in $MoS_2$ at optimal doping indeed reached the Planckian limit, another strong piece of evidence in favor of non-FL or even strange metal behavior. The results are in line with high $T_c$ superconductor and heavy Fermion superconductors[45]. Certainly the exact conclusion depends on the effective mass, however, we do not anticipate a wild change in the ratio between the measured scattering rate and Planckian scattering rate. It is not straightforward to acquire effective mass in an ionic liquid gated TMDC flake. The relatively low mobility prevents evaluating effective mass via quantum oscillations, the relatively tiny volume of the flake is not in favor of obtaining effective mass via specific heat. Meanwhile, it was technically challenging to accurately measure the optical conductivity, where scattering rate and $m^*$ can be cross-checked by the real and imaginary parts of optical conductivity, in an ionic liquid-gated device.

Hence, we believed that the observed linear *RT* at optimal doping arose from non-FL according to the feature of scattering rate. Assembling all the information, it is seen that the normal state constituted a non-FL region in the vicinity of optimal doping, which was flanked by an FL region on both sides.

## Discussions

To date, the origin of the superconducting dome in $MoS_2$ and TMDC in general remains an open question. Enhanced electron-phonon coupling due to multivalley Fermi surface evolution[5], competition between intervalley phonon scattering strength and screening[46], phonon softening associated with structural instabilities such as charge density waves[47], and quasiparticle renormalization governed by electron correlation effects[44] have been proposed as possible explanations. Most of these mechanisms tend to lead to an asymmetric superconducting dome and cannot result in a linear *RT* where the scattering rate matches the experimental observations. CDW in principle can lead to symmetric dome and linear *RT* as demonstrated in $TiSe_2$[8]. However, our transport measurements do not show explicit evidence on CDW, similar to previous reports in ionic liquid gated $MoS_2$[4, 5]. The detection of CDW in this system may need additional apparatus such as variable temperature Raman spectroscopy.

Recently, a proposal based on an extended localized-delocalized charge-Kondo breakdown transition has been successfully applied to high-$T_c$ superconductor[44]. In the model, the correlation effectively decomposes the system into a fermionic spinon band and a heavy-electron band, the local bosonic charge fluctuations behave in an effective Kondo manner and couple to these bands. It nicely reconciles the symmetric superconducting dome and strange metal behavior with a quantum critical point associated with the transition. This model inspires us to speculate that our observation is also related to a similar localized-delocalized transition and the optimal doping may correspond to the quantum critical point, by taking the localization effect induced by the ionic liquid gating into account. In our previous work[23], we proved that the ions in the ionic liquid would form a zig-zag distribution to compensate for the increasing Coulomb repulsion at large gate voltage. The zig-zag ionic lattice in turn imposes a periodic modulation on $MoS_2$ and localizes the electrons. In some regards, the ionic lattice plays a similar role as the Moiré pattern in twisted structures. The optimal doping occurs at $V_{ig}$~3.5 *V*, where the electrons are already on the verge of being strongly localized at 220 K (Fig. 1). The increased mobility due to the lowering of temperature partially restores the free-electron-like characteristics. Thereby, it is rational to think the system is indeed a mixture of localized and delocalized electrons at the optimal doping. Effective charge fluctuation arises due to the critical competition between the two flavors then drives the non-FL or even strange-metal-like behavior. The superconducting state is a result of the condensation of the quantum critical metal[44]. Below optimal doping, the system is more free-electron like; above optimal doping, it is more localized. At the moment, it seems to be challenging to theoretically handling the impact of the ionic lattice effectively without *prior* knowledge of the detailed geometric information of the ionic lattice. In principle, the geometric information may be acquired with the help of Little-Parks oscillation in the superconducting states or Aharonov-Bohm oscillation in the normal state. However, the relatively small spatial periodicity (on the order of ~1 nm, the size of molecules in the ionic liquid) of the ionic lattice requires a rather large magnetic field to observe the oscillation. Should the assertion be verified theoretically, it may suggest that *in situ* modulation of the strength of localization via ionic liquid gating could be an appealing approach in engineering exotic quantum states. Besides, it is perhaps also necessary to examine the role of the ionic liquid gating in the aspect of disorder in fermion-scalar Yukawa couplings[48].

## Conclusion

In summary, this study achieves continuous doping control of superconducting $MoS_2$ from the deeply underdoped to the deeply overdoped regime using a refined ionic liquid gating protocol, and observes a symmetric superconducting dome whose $T_c$ follows a square root dependence on carrier concentration. Moreover, we uncovered non-Fermi liquid behavior, characterized by linear temperature-resistance correspondence, anisotropy between longitudinal and transverse scattering rate, and Planckian dissipation, that is anticorrelated with superconductivity across the phase diagram. This correspondence may suggest a common physical origin for both phenomena. These findings highlight the potential of ionic liquid gating in engineering superconductivity and non-Fermi liquid behavior, and advance the understanding of the resemblance between ionic liquid-gated TMDCs and high-$T_c$ superconductors.

## Methods

**Device Fabrication and Characterization.** Thin $MoS_2$ flakes were mechanically exfoliated from bulk crystals (HQ Graphene) onto $SiO_2$ (285 nm)/Si substrates. Flakes of suitable thickness were identified via optical microscopy and confirmed by atomic force microscopy (Bruker Dimension Edge Inc.) and Raman spectroscopy.

The selected flakes were patterned into standard Hall bar geometries using a two-step electron beam lithography (EBL) process with a bilayer resist (Copolymer EL6/PMMA 495A8). The pattern was first defined by EBL (Crestec Inc.) and etched by reactive ion etching (Oxford Inc). Subsequent EBL, followed by electron-beam evaporation of Ti/Au (5/60 nm) contacts and liftoff in acetone, completed the device fabrication.

**Electrical Transport Measurements.** All electrical measurements were performed in a cryostat (TeslatronPT, Oxford Inc.). Electrical transport data were acquired using a Keithley 2634B source-measure unit for gate voltage and transfer curves, and a combination of a Keithley 6221 current source and a 2182A nanovoltmeter for other transport measurements. The use of a standard Hall bar geometry is critical for the accurate quantification of carrier concentration.

## Data availability

The data that support the findings of this study are available from the corresponding author upon reasonable request.

## Acknowledgement

This work is supported by the National Key R&D Program of China (2021YFC2202300), the National Natural Science Foundation of China (Grant No. 12204184, 12074134), Guangdong Provincial Quantum Science Strategic Initiative (GDZX2501002), Cultivation Project of Shanghai Research Center for Quantum Sciences (Grant No. LZPY2024).

## Author contributions

Q. Chen, C. Yan and S. Wang conceived and designed the experiment. Q. Chen

and H. Jian fabricated the samples. Q. Chen conducted experiments with technique supports from Y. Yan, X. Zhao, Y. Li, H. Guan, B. Gao and Zhong Wan. Q. Chen and C. Yan analyzed the data with input of D. Novko, and C. Lan. C. Yan proposed the details on analysis based on extended Kohler' rule on interpretation the data. Q. Chen and C. Yan wrote the manuscript with input of S. Wang and other authors. C. Yan and S. Wang supervised the project. All authors discussed the results and contributed to the manuscript.

## Competing interests

The authors declare no competing interests.

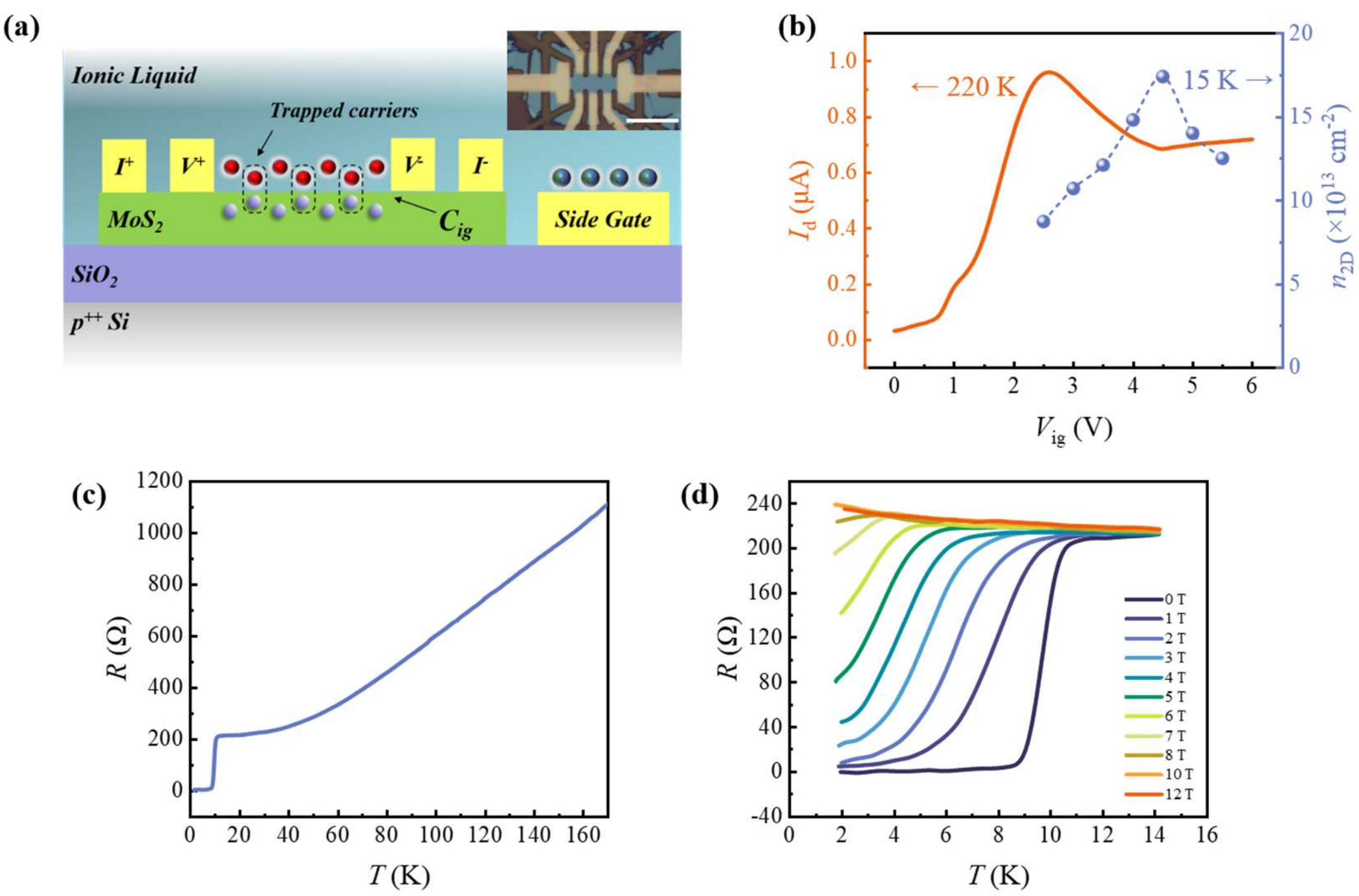

**Figure 1. Transport properties of ionic liquid-gated $MoS_2$ devices. (a)** Schematic of the device structure. The typical thickness of the studied $MoS_2$ devices is around 10 nm. Cations form a zigzag packing at the $MoS_2$ interface at large gate voltage and induce a 2D electron gas. Zigzag distributed ions create Coulomb potential traps and localize electrons. The green bar represents the $MoS_2$ crystal etched into a standard Hall bar structure. Gold squares represent Ti/Au electrodes, grey spheres representing cations. The blue transparent one is ionic liquid DEME-TFSI, with blue spheres representing anions and red spheres representing cations. The purple and grey layers represent the 285 nm $SiO_2$ and $p^{++}$ Si substrate, respectively. Inset: Optical micrograph of a typical device. Scale bar: 10 μm. **(b)** Transfer characteristic curve (left axis) and the $n_{2D}$-$V_{ig}$ curve (right axis) at 220 K

and 15 K. **(c)** $R$-$T$ curve for a typical superconducting $MoS_2$ device at $n_{2D} = 12.1\times10^{13}$ cm$^{-2}$, showing a sharp superconducting transition. **(d)** $R$-$T$ curve under out-of-plane magnetic field for the device in (c).

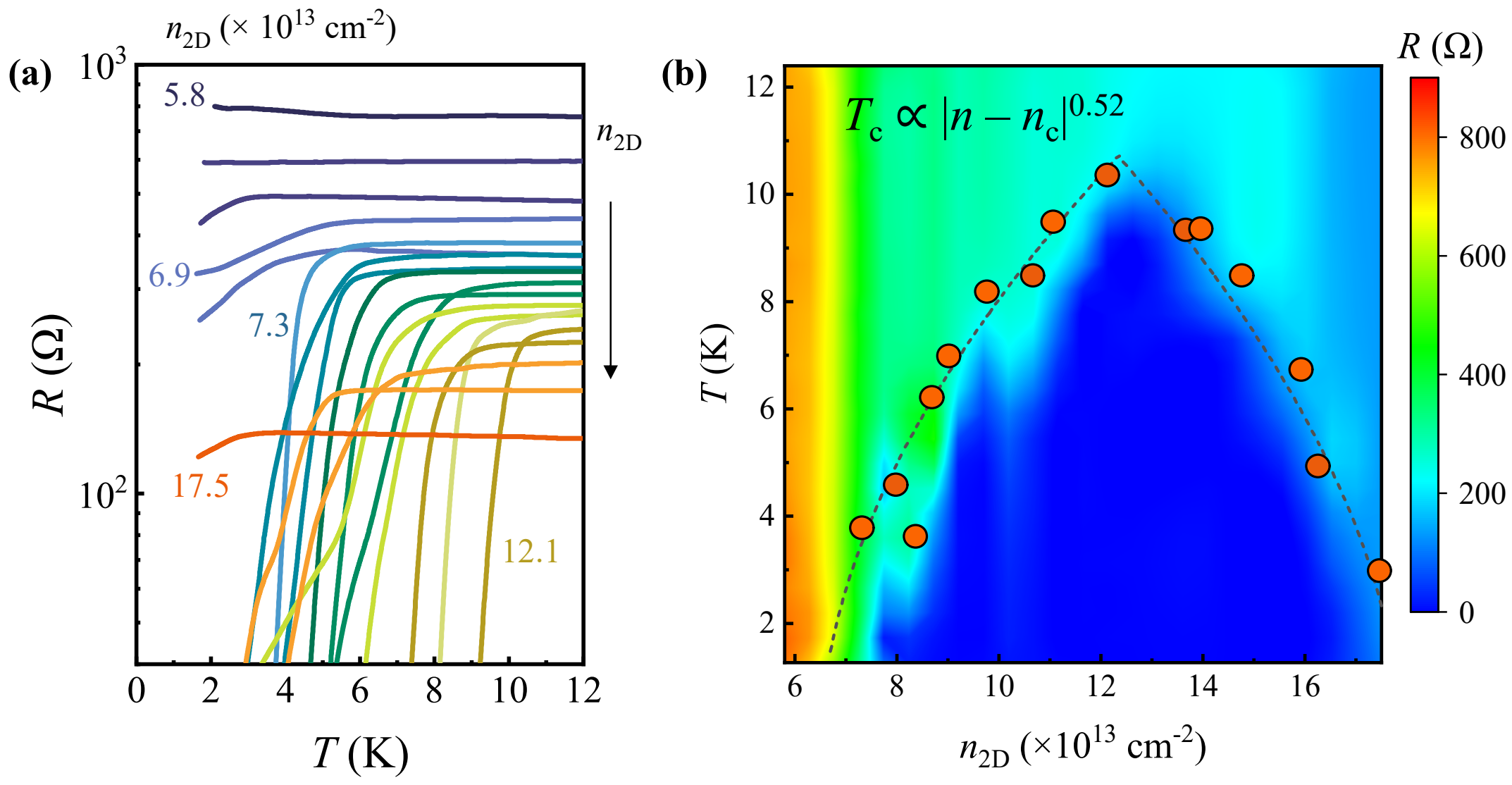

**Figure 2. Symmetric superconducting dome in ionic-$MoS_2$ devices. (a)** $R$-$T$ curves measured at different $n_{2D}$ as a function of $n_{2D}$. **(b)** Superconducting phase diagram extracted from the data in (a) by plotting the superconducting transition temperature $T_c$ (defined at 90% $R_n$) as a function of $n_{2D}$. The symmetric dome centers at the optimal doping level of $n_{2D} \sim 12.1 \times 10^{13}$ cm$^{-2}$ with a maximum $T_c \sim 10.21$ K, the dashed line fitted by $T_c \propto |n\text{-}n_c|^{0.52}$.

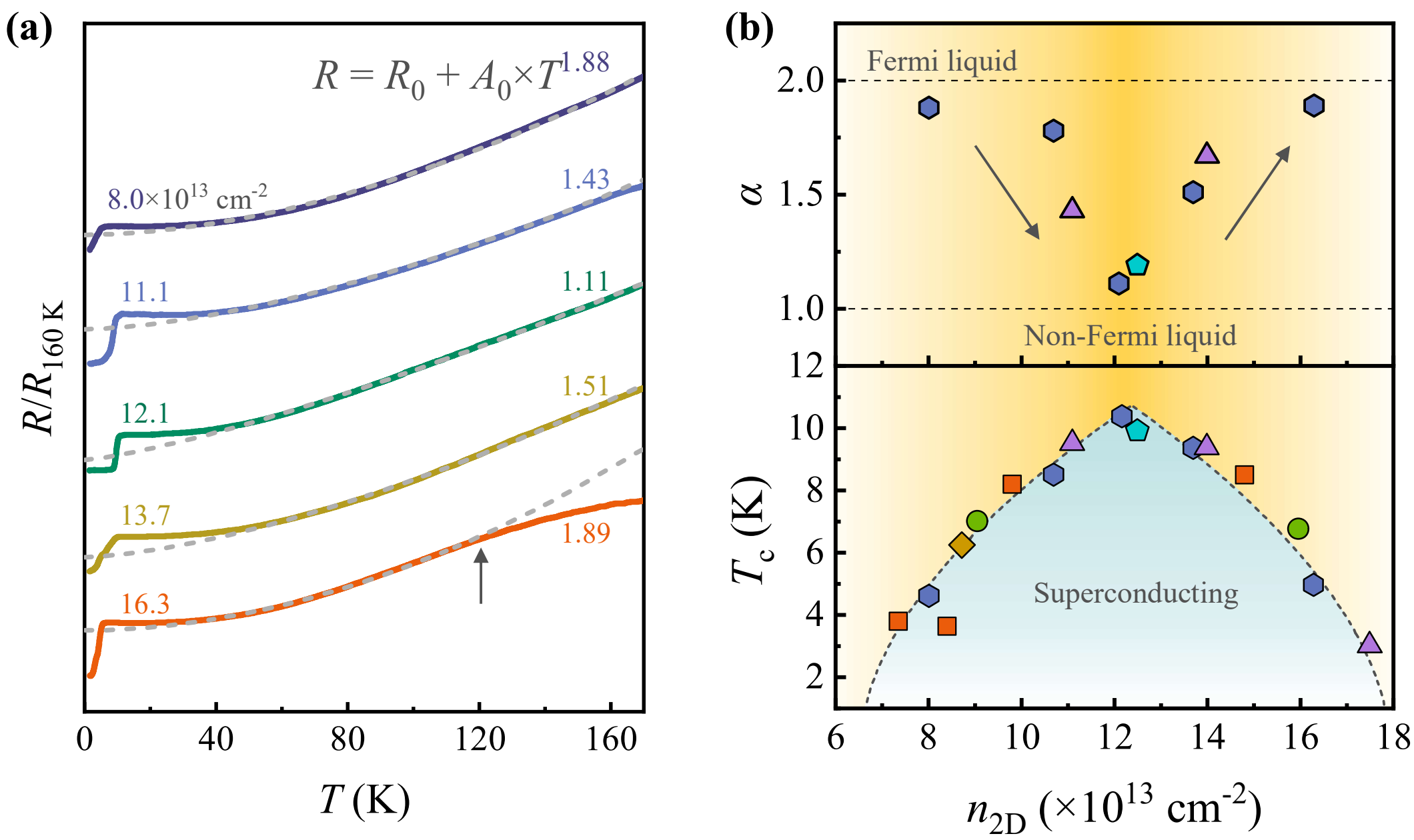

**Figure 3. Transport properties in the normal states of gated $MoS_2$ devices. (a)** $R$-$T$ curves for superconducting $MoS_2$ at various carrier concentrations $n_{2D}$, spanning the underdoped to overdoped regimes. Dashed lines are fits to the power-law form $R = R_0 + A \times T^{\alpha}$ in the high-temperature

regimes. **(b)** Doping dependence of exponent $\alpha$ (upper panel) and the superconducting transition temperature $T_c$ (lower panel). The saddle-shaped profile of $\alpha$ mirrors the superconducting dome. The markers of same color in the panel correspond to results of same devices.

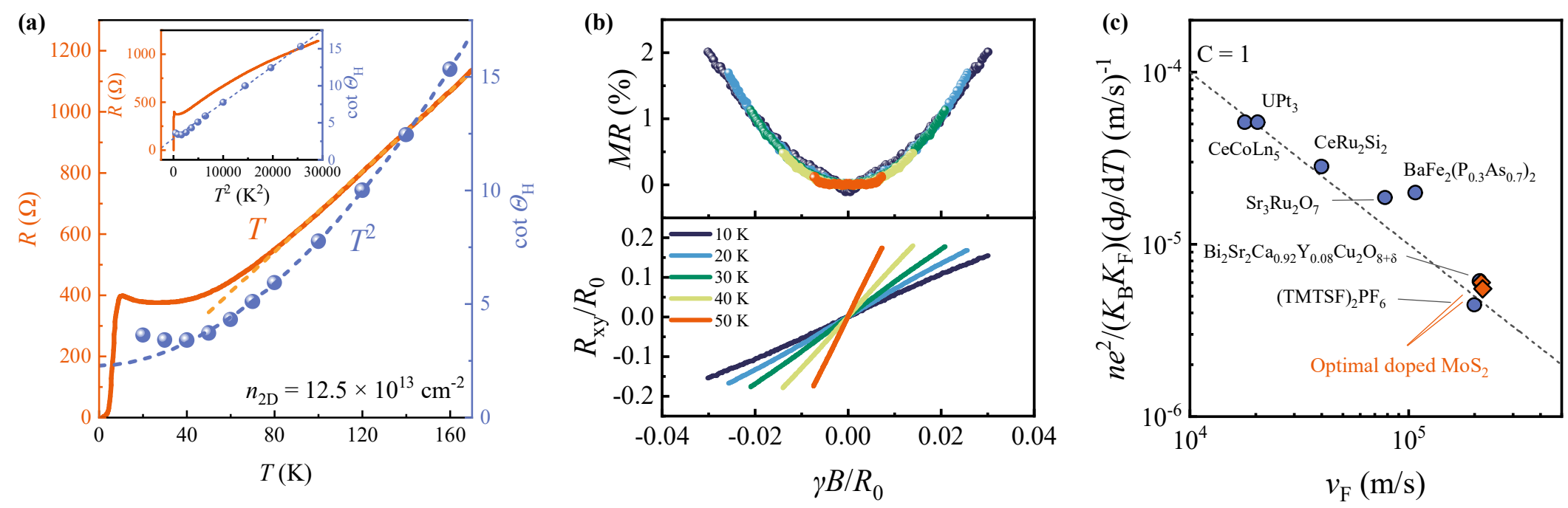

**Figure 4**. **Evidence for non-Fermi liquid transport in optimally doped $MoS_2$. (a)** Quasi-linear *R-T* curve and $T^2$-dependent Hall angle cotangent cot $\Theta_H = R_{xx}/R_{xy}$ for an optimally doped device#1. Dashed lines are fits to the data. The data indicates an anisotropy in longitudinal and transverse scattering rates, should a single parabolic band dominate the transport. Inset: The same data replotted against $T^2$. **(b)** Scaling analysis in the context of extended Kohler's rule, see main text for the details. The magnetic field is out-of-plane. Here, we scale the magnetoresistance data MR to determine the scaling factor $\gamma$, which is related to the carrier concentration and effective mass, at different temperature. The same set of $\gamma$ are then fed into the scaling analysis of Hall resistance, but fails to collapsing the data into a single curve. The analysis demonstrates an anisotropy in the scattering rates even after considering the possible contribution of multiple bands. **(c)** Ratio between the longitudinal scattering rate and Planckian limit. The horizontal axis represents the Fermi velocity $v_F$. The dashed line indicates the Planckian dissipation limit. Data for $MoS_2$ are from (a) and Supplementary Information Section 8; data for other systems are taken from Ref[45].